\documentclass[twocolumn,showpacs,showkeys,superscriptaddress]{revtex4}
\usepackage{graphicx}
\usepackage{bm}
\usepackage{color}
\usepackage{amsmath}
\usepackage{amsfonts}
\usepackage{amssymb}
\usepackage{natbib}
\usepackage{latexsym}

\begin{document}

\title{Generalized Magnetofluid Connections in Curved Spacetime}
\author{Luca Comisso}
\email{luca.comisso@columbia.edu}
\affiliation{Department of Astronomy and Columbia Astrophysics Laboratory, Columbia University, New York, NY 10027, USA}
\affiliation{Kavli Institute for Theoretical Physics, University of California, Santa Barbara, CA 93106, USA} 
\author{Felipe A. Asenjo}
\email{felipe.asenjo@uai.cl}
\affiliation{Facultad de Ingenier\'{\i}a y Ciencias, Universidad Adolfo Ib\'a\~nez, Santiago 7941169, Chile}

\begin{abstract}
The ideal magnetohydrodynamic theorem on the conservation of the magnetic connections between plasma elements is extended to non-ideal relativistic plasmas in curved spacetime. 
The existence of generalized magnetofluid connections that are preserved by the plasma dynamics is formalized by means of a covariant connection equation that includes different non-ideal effects. 
These generalized connections are constituted by 2-dimensional hypersurfaces, which are linked to an antisymmetric tensor field that unifies the electromagnetic and fluid fields. They can be interpreted in terms of generalized magnetofluid vorticity field lines by considering a $3+1$ foliation of spacetime and a time resetting projection that compensates for the loss of simultaneity between spatially separated events. The worldsheets of the generalized magnetofluid vorticity field lines play a fundamental role in the plasma dynamics by prohibiting evolutions that do not preserve the magnetofluid connectivity.

\end{abstract}

\pacs{52.27.Ny; 52.30.Cv; 95.30.Qd, 04.20.-q}
\keywords{General relativity; Relativistic plasmas; Conservation laws}

\maketitle

\section{Introduction}

A cornerstone of nonrelativistic ideal magnetohydrodynamics (MHD) is the realization that two plasma elements connected by a magnetic field line at a given time will remain connected by a magnetic field line at any subsequent time, provided that the plasma velocity field remains smooth. This property arises because a plasma that satisfies the ideal Ohm's law moves with a transport velocity that preserves the ``magnetic connections'' between plasma elements \cite{Newcomb}, a result that is generally expressed by saying that the magnetic field lines are ``frozen'' into the plasma. The importance of the magnetic field line connectivity stems from the fact that it imposes strong constraints on the plasma dynamics, in addition to providing the basis for concepts such as magnetic field line motion \cite{Newcomb}, magnetic topology \cite{Hornig97}, and magnetic reconnection \cite{Biskamp_2000}.

The above ideal MHD theorem can be cast in more clear mathematical terms \cite{Newcomb} by stating that if the ideal MHD condition ${\bm  E}+{\bm v}\times{\bm B}=0$ is satisfied, where ${{\bm E}}$ and ${\bm B}$ are the electric and magnetic field and ${\bm v}$ is the plasma velocity, then, from Faraday's equation, it follows that $d(d{\bm l}\times{\bm B})/dt=-(d{\bm l}\times{\bm B})\nabla\cdot{\bm v}-\left( (d{\bm l}\times{\bm B})\times\nabla\right)\times{\bm v}$, where $d{\bm l}$ is a vector field tangent to a curve connecting two infinitesimally close plasma elements.
Therefore, if $d{\bm l}$ is parallel to ${\bm B}$ at a given time, $d{\bm l}\times{\bm B} = 0$ remains null at all times, assuming that the velocity field remains smooth. Although not explicitly covariant, this result is valid independently of the plasma being relativistic or not. However, as discussed by Pegoraro \cite{Pego} for relativistic plasmas in the flat spacetime limit, its interpretation in terms of magnetic field lines alone requires a time resetting of the connected plasma elements in such a way to restore simultaneity when the reference frame is changed.

For general relativistic plasmas, the ideal MHD theorem on the conservation of the magnetic connections between plasma elements can be cast in a covariant form, but its interpretation in terms of magnetic field lines alone requires a specific $3+1$ foliation of spacetime \cite{asenjocomissoConnGR}, in addition to the aforementioned time resetting \cite{Pego}.   Given the ideal Ohm's law
\begin{equation}\label{Ohm1}
U^\nu F_{\mu\nu}=0\, ,
\end{equation}
where ${U}^\nu$ is the fluid 4-velocity field and $F_{\mu \nu} $ is the electromagnetic field tensor, the essential equation underlying the ideal MHD connection theorem in a general curved spacetime takes the form \cite{asenjocomissoConnGR} 
\begin{eqnarray} \label{connection_eq_ideal_MHD}
{U}^\alpha \nabla_\alpha  \left( d l^\mu F_{\mu \nu} \right) =  -  \left(\nabla_\nu {U}^\alpha\right) \left(d{l^\mu} F_{\mu \alpha}\right) \, .
\end{eqnarray} 
Here, $d l^\mu$ is the infinitesimal 4-vector separating two different fluid elements, and $\nabla_\nu$ denotes the covariant derivative associated with the spacetime metric $g_{\mu\nu}$ having signature $(-,+,+,+)$. From Eq.~\eqref{connection_eq_ideal_MHD} it follows that if $d{l^\mu} F_{\mu \alpha} = 0$ initially, then $d{l^\mu} F_{\mu \alpha}$ will remain null at all times (regularity properties of the fluid 4-velocity field are assumed). For a properly defined magnetic 4-vector $B^\mu$ [see Eq.~\eqref{E_B_Fields} below], the 4-vector event separation $d{l^\mu}$ lies in the 2-dimensional hypersurface formed by the 4-vectors $B^\mu$ and $U^\mu$. This implies that the 2-dimensional hypersurfaces generated by the 4-vectors $B^\mu$ and $U^\mu$ are preserved during the evolution of the system \cite{asenjocomissoConnGR,Carter79,Uchida97,GrallaJacob,pegoJ}. The magnetic field lines connections are finally recovered in a chosen reference frame when taking sections of these hypersurfaces at a fixed time.

Although the ideal MHD condition $U^\nu F_{\mu\nu}=0$ provides a simple and effective constitutive relation for describing the large scale and low frequency plasma dynamics, it fails to describe phenomena that are allowed by non-vanishing microscopic effects that can couple to the macroscopic plasma dynamics. Magnetic reconnection is a typical example of such multiscale coupling, where the local violation of the ideal MHD condition couple with the large-scale plasma motions. There are different non-ideal effects that can play a decisive role in the plasma dynamics, such as collisional, thermal-inertial, thermal electromotive, Hall, or current inertia effects. In these cases, Eq.~\eqref{connection_eq_ideal_MHD} is no longer valid and the aforementioned magnetic connections can be broken. However, for the flat spacetime limit, it was shown in Refs. \cite{AsComPRLconne,AsComPRLconne2} that more general magnetofluid connections can persist even in non-ideal relativistic plasmas, provided that the evolution of the system is fast compared to the dissipation timescale. These generalized connections set important constraints on the plasma dynamics by forbidding transitions between configurations with different connectivity. Therefore, in this work we aim to derive the generalized form of such invariants in curved spacetime.

We show that dynamically preserved magnetofluid connections persist also in non-ideal plasmas in curved spacetime.
In a generalized model that includes thermal-inertial effects, as well as thermal electromotive effects and Hall effects, these generalized connections can be understood through the emergence of an antisymmetric tensor field that unifies electromagnetic and fluid fields. We obtain this proof in steps, by first considering a Hall MHD plasma that encompasses thermal electromotive effects, and then adding the thermal-inertial effects of the electrons (and positrons for the pair plasma case). 
Throughout this paper, we adopt units such that the speed of light is $c = 1$.

\section{Generalized Ohm's law in Curved Spacetime}

In what follows, we consider a plasma consisting of two fluids, where one fluid is composed of positively charged particles with mass $m_+$ and electric charge $e$, and the other fluid consists of negatively charged particles with mass $m_-$ and electric charge $-e$. From the two-fluid equations in curved spacetime, one can then derive a generalized Ohm's law retaining two-fluid effects that are neglected in general relativistic MHD \cite{lich,anile}. As usual, we define the average and difference variables as follows:
\begin{equation}
n = \frac{{m_+}{n_+}{\gamma_+ '} + {m_-}{n_-}{\gamma_- '}}{{m_+} + {m_-}} \, ,
\end{equation}
\begin{equation}
p = p_+ + p_- \,  ,   \quad  \Delta p = p_+ - p_-   \, ,
\end{equation}
\begin{equation}
h = {n^2} \left( {\frac{{{h_ + }}}{{n_ + ^2}} + \frac{{{h_ - }}}{{n_ - ^2}}} \right)   \, ,
\end{equation}
\begin{equation}
{U^\mu } = \frac{{{m_ + }{n_ + }U_ + ^\mu  + {m_ - }{n_ - }U_ - ^\mu }}{{n({m_ + } + {m_ - })}}   \, ,
\end{equation}
\begin{equation}
{J^\mu } = e({n_ + }U_ + ^\mu  - {n_ - }U_ - ^\mu )   \, ,
\end{equation}
where $\gamma_\pm '$ is the Lorentz factor of the two fluids observed by the local center-of-mass frame of the plasma, $n_\pm$, $p_\pm$, and $h_\pm$ are the proper particle number density, proper pressure, and relativistic enthalpy density of the two fluids, respectively, and $U_\pm^\mu$  is the 4-velocity (subscripts $+$ and $-$ are used to indicate the positively and negatively charged fluids). The 4-velocity $U^\mu$ is normalized as $U^\mu U_\mu=g_{\mu\nu} U^\mu U^\nu=-1$. Finally, $J^\mu$ is the 4-current density.

Adopting these one-fluid variables, one can express the generalized Ohm's law in curved spacetime as \cite{Koide2010,comiAsenjblackhole}
\begin{eqnarray}\label{GeneralOhmlaw}
&&\frac{1}{4en}\nabla_\nu\left[\frac{\xi h}{ne}\left(U^\mu J^\nu + J^\mu U^\nu -\frac{\Delta\mu}{ne}J^\mu J^\nu\right) \right]\nonumber\\
&&=\frac{1}{2ne}\nabla^\mu\left(p\Delta\mu-\Delta p\right)+\left(U_\nu-\frac{\Delta\mu}{ne} J_\nu\right)F^{\mu\nu} \nonumber\\
&&-\eta \left[ {J^\mu - \rho '_e  (1 + \Theta) U^\mu} \right] \, ,
\end{eqnarray}
where $\nabla^\mu=g^{\mu\nu}\nabla_\nu$. Here, $\rho '_e = -U_\nu J^\nu$ is the charge density observed by the local center-of-mass frame, $\Theta$ is the thermal energy exchange rate between the two fluids \cite{Koide2010}, and $\eta$ is the electrical resistivity, which is considered as a phenomenological parameter in this model. 
In Eq.~\eqref{GeneralOhmlaw}, we have also introduced the variables
\begin{equation}
\Delta\mu=\frac{m_+-m_-}{m_+ + m_-} \, 
\end{equation}
and 
\begin{equation}
\xi=1-(\Delta\mu)^2 = \frac{4m_+m_-}{(m_++m_-)^2}  \, .
\end{equation}
Therefore, for an electron-ion plasma we have $\xi\approx 4m_-/m_+$, while we have $\xi=1$ for a pair plasma.

The derivation of the generalized Ohm's law \eqref{GeneralOhmlaw} assumes ${n_+} \approx {n_-}$ and $\Delta h \ll h$, where $\Delta h = m{n^2}({h_ + }/{m_ + }n_ + ^2 - {h_-}/{m_-}n_-^2)/2$ is the difference between the enthalpy densities of the fluids. We can clearly recognize several terms that are neglected in standard MHD. The terms in the left-hand side of Eq.~\eqref{GeneralOhmlaw} are due to the inertia effects of the electric current. They enter through the electron rest mass $m_-$ and depend also on the temperature via $h_\pm = m_\pm n_\pm {{{K_3}(m_\pm/{k_B}T_\pm)}}/{{{K_2}(m_\pm/{k_B}T_\pm)}}$ \cite{Chandra1938,Synge1957}, where $K_2$ and $K_3$ are the modified Bessel functions of the second kind of orders two and three,  $T_\pm$ are the temperatures of each fluid, and $k_B$ is the Boltzmann constant. The terms with respect to pressure gradients in the right-hand side of Eq.~\eqref{GeneralOhmlaw} represent the thermal electromotive force. In an electron-ion plasma ($m \approx m_+ \gg m_-$), the thermal electromotive force affects the plasma dynamics essentially through the electron pressure $p_-$, while in a pair plasma ($m_+ = m_-$) it comes into play through the pressure difference $\Delta p$. The fourth term in the right-hand side of Eq. \eqref{GeneralOhmlaw} takes into account the contribution of the Hall effect. This term vanishes in the pair plasma case ($\Delta\mu = 0$). 
Finally, the terms proportional to the resistivity $\eta$ represent the frictional 4-force density between the fluids.

If the evolution of the system is fast compared to the dissipation timescale, the frictional force between the fluids can be neglected. This is the case under typical astrophysical conditions where the plasma is essentially collisionless. In this case, we can identify important topological invariants that constrain the plasma dynamics in curved spacetime by forbidding transitions between different topological configurations. We analyze these topological invariants in the next two sections.

\section{Preserved Connections in Hall MHD plasmas} \label{SecHall}

In this section, we focus on collisionless Hall MHD plasmas.  Therefore, we assume that the thermal-inertial terms are much smaller than the Hall term, which is retained in Ohm's law. Furthermore, if the length scale $L_p$ of plasma pressure variations is such that $(m_-/m_+)(L_p/L_{th})(h/ne)\ll p$, where $L_{th}$ is the length scale associated with the thermal-inertial effects, also pressure gradients terms should be retained. Thus, the generalized Ohm's law in curved spacetime reduces to
\begin{equation}\label{GeneralOhmlawHall}
{\cal U}_\nu F^{\mu\nu} + \frac{{\nabla^\mu}\chi}{ne} =0\, ,
\end{equation}
where
\begin{equation}\label{generalvelocitycurrent}
{\cal U}_\nu = U_\nu - \frac{\Delta \mu}{ne} J_\nu\, 
\end{equation}
is a generalized transport 4-velocity, which takes into account the Hall effect, while 
\begin{equation}\label{chi_def}
\chi= \frac{p\Delta\mu-\Delta p}{2} \, 
\end{equation}
encapsulates the thermal contributions to this reduced Ohm's law. 
Equations \eqref{GeneralOhmlawHall}-\eqref{chi_def} are general to collisionless Hall MHD plasmas in curved spacetime. It can be shown that under the assumption $n_- \approx n_+$, the transport 4-velocity is ${\cal U}^\mu \approx(n_+/n)\left[ (1-m_-/m_+)U_-^\mu +(m_-/m_+)U_+^\mu\right]$, where $n/n_+\approx(1-m_-/m_+){\gamma_+ '} +(m_-/m_+) {\gamma_- '}$.  Then,  for $m_-/m_+ \rightarrow 0$ and immobile ions, the transport 4-velocity becomes ${\cal U}^\mu = U_-^\mu$, namely the electron fluid 4-velocity.

The Ohm's law \eqref{GeneralOhmlawHall} yields a connection theorem similar to the one pertaining to ideal MHD in curved spacetime \cite{asenjocomissoConnGR,Carter79,Uchida97,GrallaJacob,Novikov73,BekensteinOron78,CarterAGN79}. Indeed, we can show that Eq. \eqref{GeneralOhmlawHall} implies that the electromagnetic field is Lie dragged by the velocity field ${\cal U}_\nu$, and that the 2-dimensional connection hypersurfaces generated by the magnetic 4-vector $B^\mu$ and the transport 4-velocity ${\cal U}^\mu$ are preserved during the evolution of the system. To this purpose, we consider the convective derivative ${\cal U}^\nu \nabla_\nu$ defined along the generalized 4-velocity given by Eq. \eqref{generalvelocitycurrent}. It is straightforward to show that the electromagnetic potential ${A_\mu}$ is convected as
\begin{equation} \label{Ohm2}
{\cal U}^\nu \nabla_\nu  {A_\mu}  = {\cal U}^\nu \nabla_\mu A_\nu + \frac{{\nabla_\mu}\chi}{ne} \, ,
\end{equation}
which follows from the covariant form of the electromagnetic field tensor
$F_{\mu\nu}=\nabla_\mu A_\nu-\nabla_\nu A_\mu=\partial_\mu A_\nu-\partial_\nu A_\mu$.
Then, we can write the convective derivative of the electromagnetic field tensor as
\begin{eqnarray}\label{convF1}
{\cal U}^\alpha \nabla_\alpha F_{\mu \nu}&=& {\cal U}^\alpha \nabla_\alpha (\nabla_\mu A_\nu - \nabla_\nu A_\mu) \nonumber\\
&=& {\cal U}^\alpha {\nabla _\mu }{\nabla _\alpha }{A_\nu } -{\cal U}^\alpha  {\nabla _\nu }{\nabla _\alpha }{A_\mu } \nonumber\\
&&+ {\cal U}^\alpha R_{\beta \nu \mu \alpha} A^\beta - {\cal U}^\alpha R_{\beta \mu \nu \alpha} A^\beta  \, ,
\end{eqnarray}
where we have exploited the noncommutative properties of the covariant derivatives.
Applying the Bianchi identity for the Riemann curvature tensor, $R_{\beta \nu \mu \alpha}  +  R_{\beta \mu \alpha \nu} + R_{\beta \alpha \nu \mu} = 0$, together with Eq. \eqref{Ohm2}, we obtain
\begin{eqnarray}\label{convF3}
{\cal U}^\alpha \nabla_\alpha F_{\mu \nu}&=&  \left({\nabla_\nu} {\cal U}^\alpha\right) F_{\alpha \mu}  - \left(\nabla_\mu {\cal U}^\alpha\right) F_{\alpha \nu}+{\cal U}^\alpha \left(\nabla_\mu \nabla_\nu A_\alpha \right)\nonumber\\
&& - {\cal U}^\alpha \left(\nabla_\nu \nabla_\mu A_\alpha\right)  - {\cal U}^\alpha  R_{\beta \alpha \nu \mu}  A^\beta\nonumber\\
&&+\nabla_\mu (1/ne)\nabla_\nu\chi-\nabla_\nu (1/ne)\nabla_\mu\chi  \, .
\end{eqnarray}
If we assume that an equation of state of the form $p_\pm = p_\pm (n_\pm)$ holds, then $\nabla_\mu (1/ne)\nabla_\nu\chi-\nabla_\nu (1/ne)\nabla_\mu\chi = 0$. Finally, exploiting the noncommutative properties of the convective derivatives, we end up with
\begin{equation} \label{eq_F}
{\cal U}^\alpha \nabla_\alpha F_{\mu \nu}  = \left(\nabla_\mu {\cal U}^\alpha\right) F_{\nu \alpha}   - \left(\nabla_\nu {\cal U}^\alpha\right) F_{\mu \alpha} \, ,
\end{equation}
implying that the electromagnetic field $F_{\mu\nu}$ is Lie-dragged with the 4-velocity ${\cal U}^\mu$ given by Eq.~\eqref{generalvelocitycurrent}, instead of the fluid 4-velocity ${U}^\mu$ that characterizes the ideal MHD limit \cite{lich,acht83,Uchida97,GrallaJacob,asenjocomissoConnGR}.

In order to prove the conservation of magnetic connections in general relativistic Hall MHD plasmas, let us consider a spacelike event-separation 4-vector $d l^\mu = x'^{\mu} - x^{\mu}$ transported by the 4-velocity ${\cal U}^\mu$. Simultaneous events are defined by the vanishing component $dl^0=0$ \cite{asenjocomissoConnGR,Pego}.
We take the convective derivative of $d l^\mu$ along the 4-velocity ${\cal U}^\mu$,  which gives 
\begin{eqnarray} \label{convsep1}
{\cal U}^\nu \nabla_\nu d l^\mu  &=& {\cal U}^\nu \nabla_\nu x'^{\mu} -  {\cal U}^\nu \nabla_\nu x^{\mu}   \nonumber\\
&=& {\cal U}^\nu  \partial_\nu  x'^{\mu} - {\cal U}^\nu  \partial_\nu  x^{\mu} + {\cal U}^\nu  {\Gamma^\mu}_{\nu \lambda} x'^{\lambda}  -  {\cal U}^\nu  {\Gamma^\mu}_{\nu  \lambda} x^\lambda   \nonumber\\
&=&  {\cal U}'^{\mu} - {\cal U}^{\mu} + {\cal U}^\nu {\Gamma^\mu}_{\nu \lambda} d{l^\lambda}\nonumber\\
&=& d {l^\lambda} \nabla_\lambda {\cal U}^{\mu}  \, ,
\end{eqnarray}
where ${\Gamma^\mu}_{\nu \lambda}$ are the Christoffel symbols associated with the metric $g_{\mu\nu}$. Then, we can duly calculate the convective derivative of the quantity $d l^\mu F_{\mu \nu}$ by using Eqs.~\eqref{eq_F} and \eqref{convsep1}, which leads us to the connection equation  
\begin{eqnarray} \label{connection_eq}
{\cal U}^\alpha \nabla_\alpha  \left( d l^\mu F_{\mu \nu} \right) &=&  d{l^\alpha} \left(\nabla _\alpha {\cal U}^\mu\right)  F_{\mu \nu} \nonumber\\
&&+  d{l^\mu}\left( \nabla_\mu {\cal U}^\alpha F_{\nu \alpha}    -  \nabla_\nu {\cal U}^\alpha F_{\mu\alpha} \right) \nonumber\\
&=&  -  \left(\nabla_\nu {\cal U}^\alpha\right) \left(d{l^\mu} F_{\mu \alpha}\right) \, .
\end{eqnarray}
Equation \eqref{connection_eq} shows that if initially we have 
\begin{equation}\label{condFroz}
d{l^\mu} F_{\mu \alpha} = 0 \, ,
\end{equation}
and the 4-velocity field ${\cal U}^\nu$ remains smooth, then ${\cal U}^\alpha \nabla_\alpha (d l^\mu F_{\mu \nu}) = 0$ at every time, implying that $d{l^\mu} F_{\mu \alpha}$ will remain null at all times. Therefore, the only difference with respect to the ideal MHD case is given by the fact that $d{l^\mu} F_{\mu \alpha} = 0$ is preserved by means of the 4-velocity field ${\cal U}^\nu$ instead of $U^\nu$.

As discussed in Ref. \cite{asenjocomissoConnGR}, to specify the connection concept in curved spacetime in terms of magnetic field line connections, we need to analyze Eqs.~\eqref{connection_eq} and \eqref{condFroz} in the $3+1$ formalism (e.g., Ref. \cite{TM_82,Thorne86}), where the 4-dimensional spacetime is foliated by 3-dimensional spatial hypersurfaces of constant coordinate time. To this purpose, we write the line element in the Arnowitt-Deser-Misner (ADM) form \cite{ADM62,misner}
\begin{eqnarray} \label{spacetime_metric}
ds^2&=&g_{\mu\nu}dx^\mu dx^\nu\nonumber\\
&=&-\alpha^2 dt^2+\gamma_{ij}\left(dx^i+\beta^i dt\right)\left(dx^j+\beta^j dt\right)\, ,
\end{eqnarray}
with latin indices running for spatial components. Here, $\alpha$ is the lapse function, $\beta^ \mu=(0,\beta^i)$ is the shift vector, and $\gamma_{ij}$ is the 3-metric tensor on the spacelike hypersurfaces $\Sigma_t$ of constant time $t$. The timelike unit vector field normal to $\Sigma_t$ is given by a normalized timelike vector field $n^\mu$ with the form 
\begin{eqnarray} \label{}
    && n_\mu = - \alpha {\nabla_\mu}t = (-\alpha,0_i) \, , \nonumber \\
    && n^\mu=(1/\alpha, -\beta^i/\alpha) \, ,
\end{eqnarray}
which fullfill the normalization condition $n_\mu n^\mu=-1$. The induced metric on the spacelike hypersurface $\Sigma_t$ is
\begin{equation} \label{spacetime_metric}
{\gamma _{\mu \nu }} = {g_{\mu \nu }} + {n_\mu }{n_\nu } \, ,
\end{equation}
which satisfies the orthogonality condition $n^\mu\gamma_{\mu\nu}=0$.
The hypersurfaces $\Sigma_t$ can be viewed as an absolute space at different instances of time $t$, while the 4-vector $n^\mu$ can be interpreted as the 4-velocity of the local fiducial observer (FIDO) at rest in this absolute space.

Using the $3+1$ split of spacetime, the electromagnetic field tensor is decomposed as
\begin{equation}\label{emField_n}
{F^{\mu \nu }} = {E^\mu }{n^\nu } - {E^\nu }{n^\mu } - {\epsilon ^{\mu \nu \rho \sigma}}{B_\rho }{n_\sigma } \, ,
\end{equation}
where ${\epsilon ^{\mu \nu \rho \sigma}} =  [\mu \nu \rho \sigma]/\sqrt { - g} $, with $[\mu \nu \rho \sigma]$ indicating the fully antisymmetric symbol and $g = \det {g_{\mu \nu }}$.
The electric $E^\mu$ and magnetic $B^\mu$ 4-vectors measured in the FIDO frame are
\begin{equation} \label{E_B_Fields}
E^\mu = {n_\nu }{F^{\mu \nu }} \, , \qquad
B^\mu = \frac{1}{2}{n_\rho }{\epsilon ^{\rho \mu \sigma \tau }}{F_{\sigma \tau }}  \, .
\end{equation}
In this description, both fields are purely spatial, whereby ${n_\mu }{E^\mu } = 0$ and ${n_\mu }{B^\mu } = 0$.

We can now determine the connection condition by substituting Eq.~\eqref{emField_n} into Eq.~\eqref{condFroz}. This gives us
\begin{equation}\label{condF3mas1}
n_\mu \left(d{l^\nu} E_{\nu}\right)-\epsilon_{\mu\nu\rho\sigma}dl^\nu B^\rho n^\sigma = 0\, ,
\end{equation}
where we have used that $dl^\mu n_\mu = 0$ for simultaneous events ($dl^0=0$).  
This simultaneity condition does not affect the generality of the analysis presented here, since if $dl^0 \neq 0$, one can always restore simultaneity between spatially separated events by performing the transformation \cite{Pego} $dl^\mu \rightarrow dl'^\mu = dl^\mu + {\cal U}^\mu d\lambda$, with $\lambda$ indicating a scalar function, such that simultaneity can be realized with $dl'^0 = 0$. Indeed, due to the validity of Ohm's law \eqref{GeneralOhmlawHall}, this transformation leaves the connection equation \eqref{connection_eq} unaltered.
We then project Eq.~\eqref{condF3mas1} along the hypersurface-orthogonal (temporal) direction by contracting it with $n^\mu$, which gives $d{l^\nu} E_{\nu}= 0$, showing that the electric field is orthogonal to the event-separation 4-vector. On the other hand, by projecting Eq.~\eqref{condF3mas1} into the hypersurface-tangential (spatial) slice through the projector tensor ${\gamma^\mu}_\nu = {\delta^\mu}_\nu + n^\mu n_\nu$, we have
\begin{equation}\label{condF3mas1Space}
\epsilon_{0ijk}dl^j B^k = 0\, .
\end{equation}
Therefore, in the $3+1$ formalism, and under the simultaneity condition $dl^0 = 0$ according to our choice of the spacetime foliation, the condition \eqref{condFroz} is equivalent to the vectorial condition $d{\bm{l}} \times {\bm{B}} = 0$ and comprise the condition $d{\bm{l}} \cdot  {\bm{E}} = 0$. According to the connection equation \eqref{connection_eq}, this implies that if $d{\bm{l}} \times {\bm{B}}$ vanishes initially, i.e. the vector field tangent to a curve connecting two fluid elements, $d{\bm{l}}$, is aligned with the magnetic field ${\bm{B}}$, the plasma evolution is such that this property remains preserved over time. Hence, an Ohm's law of the form \eqref{GeneralOhmlawHall} for Hall MHD plasmas in curved spacetime does not allow the breaking of magnetic connections between fluid elements, essentially electron fluid elements. Magnetic reconnection can occur if Eq. \eqref{GeneralOhmlawHall} becomes invalid, as it happens if $\Delta h \gg h$ \cite{Kawa17}, or if resistivity is non-negligible.

The condition \eqref{condFroz} holds also if $dl^0 \neq 0$. In this case,  the 4-vector event separation $dl^\mu$ remains in the 2-dimensional hypersurface generated by the 4-vectors ${\cal U}^\mu$ and $B^\mu$. Therefore, the connected field lines are generalized in a covariant way by using worldsheets of the magnetic field lines \cite{asenjocomissoConnGR,Carter79,Uchida97,GrallaJacob}. On the other hand, differently from the ideal Ohm's law case, the worldsheets of the magnetic field lines are advected by the 4-velocity ${\cal U}^\mu$ instead of the plasma 4-velocity $U^\mu$.

\section{Preserved Connections in Extended MHD plasmas}

In Sec. \ref{SecHall}, we neglected all the electron inertia terms under the assumption that they are much smaller than the Hall term and/or the pressure gradients terms. However, when electron current layers become important, these terms cannot be neglected as they play a crucial role in the plasma dynamics. Indeed, the finite inertia of the electrons 
can break the magnetic connections and magnetic reconnection can take place \cite{Biskamp_2000}. Nevertheless, magnetic reconnection mediated by electron inertia preserves other generalized field lines connections. This property, which is well known in the nonrelativistic regime \cite[e.g.][]{OP_1993,CafGrasso98,Pegoraro2004,Comisso2013,Lingam16}, has been shown to hold in special relativity \cite{AsComPRLconne,AsComPRLconne2,Pegoraro15} with a proper definition of generalized electromagnetic fields. Here we show that generalized connections can also exist in curved spacetime, and that they can be interpreted in terms of generalized magnetofluid vortex lines when a $3 + 1$ split of spacetime is employed.

For collisionless plasmas ($\eta=0$), we can rewrite the generalized Ohm's law \eqref{GeneralOhmlaw} as
\begin{eqnarray}\label{GeneralOhmlaw2}
\frac{J^\nu}{ne} \nabla_\nu\left(\frac{\xi h}{4n e}{\cal U}^\mu \right)&+&U^\nu\nabla_ \nu\left(\frac{\xi h}{4e^2n^2} J^\mu\right)\nonumber\\
&&\qquad = \frac{{\nabla^\mu}\chi}{ne}  +{\cal U}_\nu F^{\mu\nu}\, ,
\end{eqnarray}
where we have used the continuity equations $\nabla_\mu \left(n U^\mu\right)=0$ and $\nabla_\mu J^\mu=0$. Following the procedure outlined in Refs.~\cite{AsComPRLconne,AsComPRLconne2}, we can cast the above equation in the form
\begin{equation}\label{GeneralOhmlaw3}
{\cal U}_\nu{\cal M}^{\mu\nu}+\frac{1}{n e} \nabla^\mu\chi -\nabla^\mu\left(\frac{\xi h}{4\Delta\mu\, ne}\right)=\Sigma^\mu\, ,
\end{equation}
where we have defined the generalized magnetofluid field tensor
\begin{equation}\label{generM}
{\cal M}^{\mu\nu}=F^{\mu\nu}-\frac{\xi}{4 \Delta\mu} W^{\mu\nu} \, ,
\end{equation}
with ${W}^{\mu\nu}$ indicating the antisymmetric flow field tensor
\begin{equation}\label{defW}
{W}^{\mu\nu}= \nabla^\mu\left(\frac{h}{n e}{\cal U}^\nu\right)  -\nabla^\nu\left(\frac{h}{n e}{\cal U}^\mu\right) \, .
\end{equation}
Furthermore, we have introduced the effective thermal-inertia 4-vector 
\begin{eqnarray}
\Sigma^\mu=\frac{\zeta}{n e}\nabla^\mu\left(\frac{h}{ne} \right)+\frac{\xi}{4\Delta\mu} U^\nu\nabla_\nu\left(\frac{h}{ne}U^\mu\right)\, ,
\end{eqnarray}
along with the scalar 
\begin{equation}
\zeta= \frac{\xi}{4}  \left( U_\mu J^\mu  - \frac{\Delta\mu}{2n e} J_\mu J^\mu \right)  \, .
\end{equation}

The generalized Ohm's law \eqref{GeneralOhmlaw3} includes all the thermal-inertial effects, as well as thermal electromotive effects and Hall effects. The generalized magnetofluid field tensor ${\cal M}^{\mu\nu}$ represents an effective field tensor that unifies the electromagnetic and fluid fields. The flat spacetime limits of this tensor were introduced in Refs. \cite{AsComPRLconne} and \cite{AsComPRLconne2}, for electron-ion plasmas and pair plasmas, respectively. This tensor is similar in nature (but different in structure) to the unified magnetofluid field tensors introduced in Refs. \cite{Bekenstein87,mahajanU}. 
As we show below, the effective field tensor ${\cal M}^{\mu\nu}$ is instrumental in revealing topological invariants of collisionless plasmas beyond the MHD description.

In analogy with the previous section, we look to the evolution of the effective field tensor ${\cal M}^{\mu\nu}$. It is straightforward to show that the convective derivative of ${\cal M}^{\mu\nu}$ along the generalized 4-velocity ${\cal U}^\nu$ gives 
\begin{eqnarray} \label{eq_M}
{\cal U}^\alpha \nabla_\alpha  {\cal M}_{\mu \nu}  &=& \left(\nabla_\mu {\cal U}^\alpha\right) {\cal  M}_{\nu \alpha}   - \left(\nabla_\nu {\cal U}^\alpha\right) {\cal M}_{\mu \alpha}\nonumber\\
&&+\nabla_\mu\Sigma_\nu-\nabla_\nu\Sigma_\mu \, ,
\end{eqnarray}
where we have assumed $p_\pm = p_\pm (n_\pm)$. 
In order to obtain the dynamics of the generalized connections, let us consider a spacelike event-separation 4-vector $d l^\mu = x'^{\mu} - x^{\mu}$ that is transported with a general 4-velocity 
\begin{eqnarray} \label{}
{\cal V}^\mu = {\cal U}^\mu + Z^\mu = {\cal U}^\alpha \nabla_\alpha x^\mu \, ,
\end{eqnarray}
where $Z^\mu$ is a relative 4-velocity that fulfills the equation 
\begin{eqnarray}\label{eqforV}
\nabla_\mu Z^\alpha {\cal M}_{\alpha\nu}-\nabla_\nu Z^\alpha {\cal M}_{\alpha\mu}=\nabla_\nu\Sigma_\mu-\nabla_\mu\Sigma_\nu\, .
\end{eqnarray}
The velocity field $Z^\mu$ takes into account all the thermal--inertia  information of the plasma fluid that is usually neglected in simpler models. This velocity depends on the variation of such effects, and it is used here to prove the existence of the generalized connections that take into account such effects.

Now, the connection equation can be readily derived. We take the convective derivative of $d l^\mu$ along the 4-velocity ${\cal U}^\mu$, which gives \cite{asenjocomissoConnGR}
\begin{eqnarray} \label{convsep2}
{\cal U}^\nu \nabla_\nu  d l^\mu  =  d {l^\lambda} \nabla_\lambda {\cal V}^{\mu} \, .
\end{eqnarray}
Then, using Eqs.~\eqref{eq_M} and \eqref{convsep2}, we find that the convective derivative of $dl^\mu {\cal M}_{\mu\nu}$ follows the equation
\begin{eqnarray} \label{connection_eq4}
{\cal U}^\alpha \nabla_\alpha \left( d l^\mu {\cal M}_{\mu \nu} \right) =    -  \left(\nabla_\nu {\cal V}^\alpha\right) \left(d{l^\mu} {\cal M}_{\mu \alpha}\right) \, ,
\end{eqnarray}
which represents the magnetofluid connection equation in curved spacetime.

Equation \eqref{connection_eq4} shows that if initially we have 
\begin{equation}\label{generalconnetcionthereomextended}
d l^\mu {\cal M}_{\mu \nu} = d l^\mu F_{\mu\nu} - \frac{\xi}{4 \Delta\mu} d l^\mu W_{\mu\nu}  = 0\, ,
\end{equation}
and Eq.~\eqref{eqforV} has a well-behaved solution for the 4-velocity $Z^\mu$, then $d l^\mu {\cal M}_{\mu \nu}$ will remain null at all times. This implies the existence of 2-dimensional magnetofluid hypersurfaces that preserve their topology during the plasma dynamics. In this case, the magnetofluid field tensor ${\cal M}_{\mu \nu}$ takes the role that the electromagnetic field tensor $F_{\mu \nu}$ has for the ideal MHD and Hall MHD limits. Furthermore, the general 4-velocity ${\cal V}^\mu$ takes into account also the thermal-inertial contributions to the plasma dynamics.

To gain further insight on the connections that underlie the extended MHD plasma, we rely again on the $3+1$ split of spacetime. Therefore, the generalized magnetofluid field tensor is decomposed as
\begin{equation}\label{decompogenerali}
{{\cal M}^{\mu \nu }} = {\Xi^\mu }{n^\nu } - {\Xi^\nu }{n^\mu } - {\epsilon ^{\mu \nu \rho \sigma}}{\Omega_\rho }{n_\sigma } \, .
\end{equation}
The generalized electric $\Xi^\mu$ and magnetic $\Omega^\mu$ fields are
\begin{equation} \label{E_B_FieldsGen}
\Xi^\mu = {n_\nu }{{\cal M}^{\mu \nu }} \, , \qquad
\Omega^\mu = \frac{1}{2}{n_\rho }{\epsilon ^{\rho \mu \sigma \tau }}{{\cal M}_{\sigma \tau }}  \, ,
\end{equation}
which are both spacelike (${n_\mu }{\Xi^\mu } = 0$ and ${n_\mu }{\Omega^\mu } = 0$). The generalized magnetic field $\Omega^\mu$ can also be viewed as a magnetofluid vorticity, since $\Omega^\mu$ includes the contribution of the 4-velocity ${\cal U}^\nu$ and the thermal-inertial effects, as given by
\begin{equation} \label{emFieldn2}
\Omega^\mu =  B^\mu-\frac{\xi}{4\Delta\mu}{n_\rho }{\epsilon ^{\rho \mu \sigma \tau }\nabla_\sigma\left(\frac{h}{e n}{\cal U}_\tau\right)}\, .
\end{equation}
It is the magnetofluid vorticity $\Omega^\mu$ (in place of the magnetic field $B^\mu$) that preserves the topology during the non-ideal plasma dynamics.

The field line connectivity of the magnetofluid vorticity can be shown by choosing the simultaneity condition $dl^0=0$. As explained before, if $dl^0 \neq 0$ one can restore simultaneity by moving the endpoints of the wordline connecting the two close events along their trajectories \cite{Pego}. This is obtained by performing the transformation $dl^\mu \rightarrow dl'^\mu = dl^\mu + ({\cal U}^\mu + H^\mu) d\lambda$, such that  $dl'^0 = 0$ in the adopted spacetime foliation. Indeed, this transformation leaves the connection equation \eqref{generalconnetcionthereomextended} unaltered whenever the 4-vector $H^\mu$ fulfills the  generalized Ohm's law \eqref{GeneralOhmlaw3} \cite{AsComPRLconne}. In this case,  the 4-vector $H^\mu$ can be written as $H^\mu=({1}/{n e}) N^{\nu\mu}\nabla_\nu\chi -N^{\nu\mu}\nabla_\nu\left({\xi h}/({4\Delta\mu\, ne})\right)-N^{\nu\mu}\Sigma_\nu$, where $N^{\mu\nu}$  is the inverse of ${\cal M}^{\mu\nu}$. 
Therefore, by using Eq.~\eqref{decompogenerali} in Eq.~\eqref{generalconnetcionthereomextended} and projecting it into the spatial slice through the projector tensor ${\gamma^\mu}_\nu$, we obtain
\begin{equation}\label{condFrozgenera}
\epsilon_{0ijk}dl^j {\Omega}^k = 0\, .
\end{equation}
On the other hand, the projection of Eq.~\eqref{generalconnetcionthereomextended} along the temporal direction by contracting it with $n^\mu$ gives $d{l^\nu} {\Xi}_{\nu}= 0$.
Thus, in the $3+1$ formalism and under the simultaneity assumption $dl^0 = 0$, the condition \eqref{generalconnetcionthereomextended} is equivalent to the vectorial condition $d{\bm{l}} \times {\bm{{\Omega}}} = 0$ and comprise the condition $d{\bm{l}} \cdot  {\bm{\Xi}} = 0$. According to the connection equation \eqref{connection_eq4}, this implies that if $d{\bm{l}} \times {\bm{\Omega}}$ vanishes initially, the plasma evolution is such that this property remains preserved over time.
The 4-vector event separation $dl^\mu$ remains always in the 2-dimensional hypersurface generated by the 4-vectors ${\cal V}^\mu$ and $\Omega^\mu$. Therefore, the connected field lines are generalized in a covariant way by using worldsheets of the magnetofluid vortex lines. On the other hand, differently from the ideal Ohm's law case, the worldsheets of the magnetic field lines are advected by the general 4-velocity ${\cal V}^\mu$.

Finally, we note that the collisionless pair plasma limit of the  generalized Ohm's law \eqref{GeneralOhmlaw2} yields an antisymmetric flow field tensor 
\begin{equation}\label{}
{W}^{\mu\nu} / {\Delta\mu} \rightarrow \nabla^\nu \left(\frac{h}{n^2 e^2}{J}^\mu \right)  -\nabla^\mu \left(\frac{h}{n^2 e^2}{J}^\nu \right) \, ,
\end{equation} 
and a generalized 4-velocity ${\cal U}^\mu \rightarrow U^\mu$. Therefore, also in this limit the magnetofluid field tensor ${\cal M}^{\mu\nu}$ is constituted by a combination of electromagnetic and fluid fields.

\section{Conclusions}

In this paper, we have extended the ideal MHD theorem on the ``frozen-in'' property of the magnetic field lines to non-ideal relativistic plasmas in the presence of significant gravitational fields. This is indeed important for plasmas around black holes \cite[e.g.][]{Koide2002,Koide2010,Zam,asenjoGRLuca,EHT1} or in the early Universe \cite[e.g.][]{holcomb,TajimaShibata1997,Subra,Son}. In such cases, general relativity must be taken into account in the plasma dynamics. Furthermore, local effects that are outside the large scale and low frequency description of the ideal MHD theory can drastically modify the plasma behavior \cite[e.g.][]{Koide2010,comiAsenjblackhole,Asenjo19,parfrey,Ripperda19}, as is the case with magnetic reconnection. Indeed, the violation of the magnetic connections due to magnetic reconnection couples to the macroscopic plasma dynamics and is accompanied by the rapid release of magnetic energy into thermal and kinetic energy of the plasma, with a resulting global behavior of the system that is very different from the ideal MHD predictions.

There are different non-ideal effects that can play a decisive role in the plasma dynamics, such as collisional, thermal-inertial, thermal electromotive, Hall, or current inertia effects. Nevertheless we have shown that when the evolution of the system is fast compared to the dissipation timescale, there are different topological invariants that take the place of the magnetic field in non-ideal plasmas. In an extended MHD plasma, the preserved connections are no longer related to the electromagnetic field tensor $F^{\mu\nu}$ but to an effective field tensor that unifies the electromagnetic and fluid fields. This generalized magnetofluid field tensor ${\cal M}^{\mu\nu}$ allows the extension of the ideal MHD theorem to plasmas beyond the MHD description. In the relativistic domain, this means that there are generalized magnetofluid connections organized in 2-dimensional hypersurfaces that are dynamically preserved. In the Hall MHD limit, these connections are related to the magnetic field as in the ideal MHD theory, but they are advected by the 4-velocity ${\cal U}^\mu$ instead of the plasma 4-velocity $U^\mu$.

The 2-dimensional magnetofluid connection hypersurfaces can be interpreted in terms of magnetofluid vortex lines by employing a $3+1$ foliation of spacetime into nonintersecting spacelike hypersurfaces of constant coordinate time and the time resetting procedure introduced in Ref. \cite{Pego} to account for the loss of simultaneity in different reference frames between spatially separated events. 
This defines a new generalized vorticity $\Omega^\mu = \frac{1}{2}{n_\rho }{\epsilon ^{\rho \mu \sigma \tau }}{{\cal M}_{\sigma \tau }}$ that encompasses both the magnetic and the fluid fields. 
The field lines of this generalized magnetofluid vorticity are preserved by the non-ideal plasma dynamics and extend the usual magnetic field lines, whose topology is no more preserved when thermal-inertial effects are included. Additionally, the generalized magnetofluid vorticity is advected by a general transport 4-velocity that differs from the plasma 4-velocity that characterizes the magnetic connections in the ideal MHD limit.

The conservation of the worldsheets of generalized magnetofluid vorticity field lines set important constraints on the plasma dynamics in curved spacetime by forbidding transitions between different topological configurations of the magnetofluid vorticity sheets. In nonrelativistic ideal MHD, the preservation of the magnetic field topology is directly linked to the formation of electric current sheets \cite[e.g.][]{Zhou2016}. Topological invariants in nonrelativistic extended MHD models are also thought to be responsible for the formation of small scale structures in different nonlinear plasma processes, such as in magnetic reconnection \cite[e.g.][]{CafGrasso98}. Therefore, the investigation of topological properties of the generalized magnetofluid vorticity in curved spacetime may also provide further insights into our understanding of the nonlinear plasma dynamics in strong gravitational fields. 
Moreover, by forbidding certain class of motions, the topological invariants may guarantee the stability of the resulting solutions.

There are additional non-ideal effects that might occur in relativistic plasmas, especially in very high-energy regimes. In those cases, processes such as pair creation and annihilation, radiation-reaction, and spin effects may have to be considered in the plasma description. While the study of those effects is out of the scope of this work, we observe that a similar analysis could be performed to find a covariant generalization of the preserved field connections including those terms. Some of these high-energy effects can be included in properly redefined generalized electromagnetic fields or as dissipative processes in the generalized Ohm's law \cite{Pegoraro15}. In this case, the analysis presented in this paper remains essentially unchanged, as it applies to timescales shorter than the dissipation timescale. For those effects that cannot be included in the definition of generalized electromagnetic fields in the generalized Ohm's law, or that also modify conservation laws, such as pair production, a more general calculation could be required. These directions will be pursued in future works.

\begin{acknowledgments}
F.A.A. thanks Fondecyt-Chile Grant No. 1180139. L.C. is grateful for the hospitality of the Kavli Institute for Theoretical Physics (KITP) of the University of California Santa Barbara, where part of this work was done and supported by the National Science Foundation under Grant No. NSF PHY-1748958.
\end{acknowledgments}

\end{document}